\documentclass{article}
\usepackage{arxiv}

\usepackage[utf8]{inputenc} 
\usepackage[T1]{fontenc}    
\usepackage{hyperref}       
\usepackage{url}            
\usepackage{booktabs}       
\usepackage{amsfonts}       
\usepackage{nicefrac}       
\usepackage{microtype}      
\usepackage{lipsum}

\usepackage{amsmath}
\usepackage{caption}
\usepackage{subcaption}
\usepackage{xcolor}
\usepackage{graphicx}
\usepackage[superscript]{cite}
\usepackage{sidecap}

\title{Liquid-Liquid Critical Point in Phosphorus}

\author{
 Manyi Yang\\
  Istituto Italiano di Tecnologia (IIT)\\
  Via Morego, 30, 16163 Genova GE, Italy\\
 \And
 Tarak Karmakar\\
  Istituto Italiano di Tecnologia (IIT)\\
  Via Morego, 30, 16163 Genova GE, Italy\\
  \And
  Michele Parrinello\thanks{michele.parrinello@iit.it}\\
  Istituto Italiano di Tecnologia (IIT)\\
  Via Morego, 30, 16163 Genova GE, Italy\\
}

\date{\today}

\begin{document}
\maketitle

\begin{abstract}
The study of liquid-liquid phase transition has attracted considerable attention. One interesting example of such phenomenon is phosphorus for which the existence a first-order phase transition between a low density  insulating molecular phase and a conducting polymeric phase has been experimentally established. In this paper, we model this transition by an \emph{ab-initio} quality molecular dynamics simulation and explore a large portion of the liquid section of the phase diagram. We draw the liquid-liquid coexistence curve and determine that it terminates into a second-order critical point. Close to the critical point, large coupled structure and electronic structure fluctuations are observed.
\end{abstract}


\newpage
\twocolumn
One phenomenon that has attracted much attention is that of liquid-liquid (LL) phase transitions. Most famously, the existence of such a LL transition terminating in a critical point has been suggested as a possible  explanation for the famous water anomalies \cite{poole1992phase,sciortino1997line,Xu2011,palmer2014metastable,smallenburg2015tuning,Poole2013,gartner2020signatures,Abascal2010,Debenedetti2020}. Unfortunately the location of this transition in the so called no man's land has made its experimental study challenging.
Besides water, the existence of LL transition has been discussed in a number of other systems \cite{jakse2007liquid,vasisht2011liquid,dzyabura2013evidence,lascaris2016tunable,kurita2004critical,kurita2005abundance,huvs2014existence}. However experimental investigations in these systems are also fraught with difficulties since the reported LL transitions  occur in a regime of high temperatures and pressures.  

Among the many systems for which a LL transition has been reported a most intriguing example is that of liquid phosphorus. Phosphorus is an interesting substance with several practical applications and already in the solid phase, exhibits many allotropes. Thus, it is perhaps not a surprise that also in the liquid state, different structures can be found. Katayama \emph {et al} \cite{Katayama2000,Katayama2004} have reported experimental evidence of  the existence of a first-order LL transition as signaled by an abrupt  density jump. The low-density phase (LDL) is an insulating fluid whose structural units are P$_4$ molecules while the high-density phase (HDL) is a conducting fluid whose structure has been described as polymeric. Subsequent experiments have revealed that the slope of the LL transition line in the temperature-pressure (TP) plane is negative \cite{Monaco2003}. In addition, the existence of a liquid-liquid critical point (LLCP) at $T \geq$ 2500 K has been  conjectured but so far no firm experimental  evidence has been reported. 

This obviously calls for simulations that can clarify the behavior of liquid phosphorus and in particular, the existence of a critical point. This task is made difficult by the metal to nonmetal transition that accompanies the LL transition. The change in chemical bonding renders a standard approach to the development of a realistic phosphorus force field challenging. A possible alternative is to use \emph{ab-initio} molecular dynamics simulations in which the forces are computed on-the-fly from accurate  electronic structure calculations. Typically the forces are calculated by using density functional theory (DFT) which provides a balance between computational expediency and accuracy. A number of such simulations have been reported, and they all have shown that a DFT-based approach is able to reproduce the existence of the LL transition \cite{morishita2001liquid,ghiringhelli2005phosphorus}. A somewhat indirect attempt has also been made at estimating the critical point \cite{ghiringhelli2007simulating,zhao2017anomalous}. Although these simulations are highly illuminating, their computational cost has required some compromise as to the system size, the simulation length, and has allowed only for the limited number of thermodynamics states to be explored.

A way of obtaining \emph{ab-initio} MD accuracy at a limited  computational cost was suggested some time ago by Behler and Parrinello \cite{Behler2007,Behler2016}. Their idea was to use the generalization capabilities of neural networks (NN) to express the potential energy surface. The parameters of the NN were trained on a relatively large set of DFT calculations performed on appropriately selected atomic configurations. Since Behler and Parrinello's work, much progress has been made, and this approach has become very popular \cite{behler2011neural,yao2018tensormol,behler2017first,cheng2020evidence,smith2017ani,shen2016multiscale,shen2018molecular,cheng2020evidence}. Here we shall use, as we have done in the recent past \cite{bonati2018silicon,niu2020ab,YANG2021}, the Deep MD code \cite{zhang2018deep,end-to-end}.  In the case of phosphorus, an approach similar in spirit but based on the Gaussian Kernel representation of the potential has already been used to study its allotropes and liquid state \cite{deringer2020general}. However, the exploration of the high pressure and high temperature part of the phase diagram has been limited.

In the Behler-Parrinello-type of approach, a judicious choice of the training configurations is important. In our case, the choice of configurations is complicated by the fact that a first-order phase transition is a rare event. Therefore, the transition state configurations that are essential for determining the transition barrier between one phase and the other are rarely sampled and are not encoded in the NN.  For this reason,  when studying rare events, we have proposed to train the NN extracting configurations from an enhanced sampling run where the low probability, but highly important transition state arrangements are sampled \cite{bonati2018silicon,niu2020ab,YANG2021}. Furthermore, in the study of phase diagrams, it is convenient to follow Ref. \cite{niu2020ab} and \cite{pablo-MTP-1,pablo-MTP-2} and use a multithermal multibaric (MultiTP) enhanced sampling method. The MultiTP approach allows entire regions of the TP phase diagram to be explored at a computational cost comparable to that of a single standard simulation at a given temperature and pressure. A full description of the technical details can be found in the supporting information (SI). Here we mention only the fact that we use the OPES version of the MultiTP approach \cite{piaggi2019multithermal,invernizzi2020unified}. OPES, like many other enhanced sampling methods \cite{laio2002escaping,barducci2008well,valsson2016enhancing}, relies on the definition of a collective variable (CV) whose fluctuations are enhanced by the method \cite{invernizzi2020rethinking,invernizzi2020unified}. A natural choice would have been to use the density as a CV that is the natural order parameter for the phase transition. However since this CV performed poorly, we choose instead the value of the first peak of the Debye structure factor (see Fig. S2 in SI)  \cite{niu2018molecular}. However, most of the time we express our results as a function of the density that is a standard thermodynamic variable.  

The strategy for generating the NN training configurations follows the active learning scheme of Ref. \cite{zhang2019active,zhang2020dp,YANG2021}. That is, one first performs a number of standard \emph{ab initio} MD calculations in a grid of thermodynamic conditions. These configurations are used to obtain a first guess of the NN potential. In reality, four slightly different NN models are obtained by starting the NN stochastic optimization of the NN from different initial conditions. One of these models is then used to drive the MultiTP-OPES calculation.   
Periodically we check whether all four NN models predict similar forces. If the discrepancy between their predictions exceeds a preassigned threshold (see SI), we calculate the DFT energies and forces for those particular configurations and add these data to the training set and at a fixed interval re-train all four NN models. This procedure is continued until no significant discrepancy between the four models is observed. The calculations presented here are based on the SCAN+D3 exchange-correlation model that gives a good agreement between experiment and theory (see Fig. S6 in SI). 

\begin{figure}[!h]
  \centering
  \includegraphics[width=0.48\textwidth]{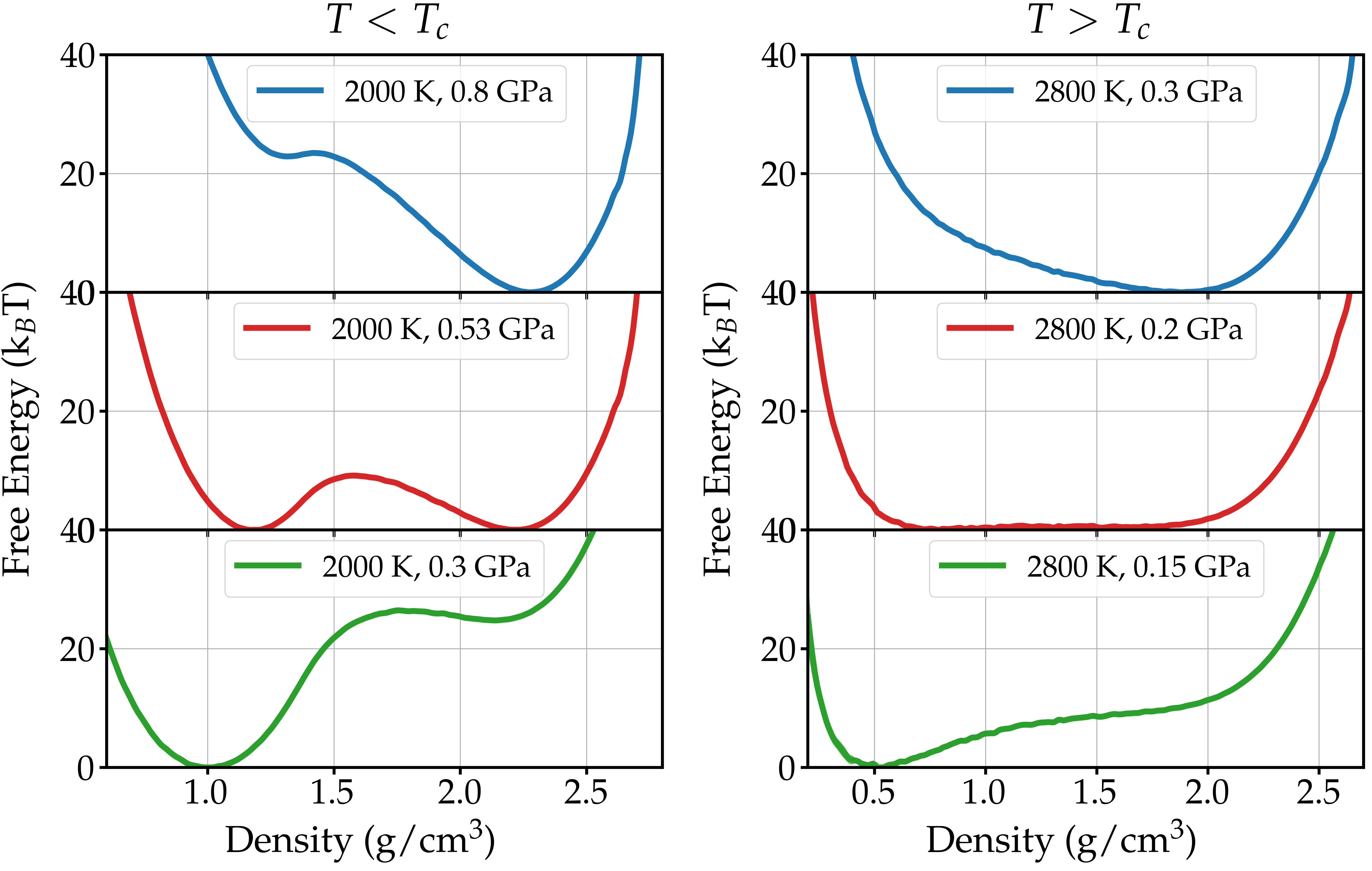}
  \caption{Free-energy surfaces (FESs) as a function of the density for  $T=$ 2000 K $(T < Tc)$ (left) and $T=$ 2800 K $(T > Tc)$ (right). For each $T$, three FESs at different $P$ values are shown. Error bars, calculated using the weighted block-average technique (number of blocks = 4) discussed in Ref. \citenum{invernizzi2020unified}, are smaller than the linewidth.}
  \label{fgr:FES}
\end{figure}

Since our MD calculations are  based on a MultiTP approach, we can calculate the system properties in the range of temperatures and pressures that is specified at the beginning of the calculation, without having to perform new simulations. Thus, it is painless to identify the transition line and other significant thermodynamic points. The procedure used to draw the coexistence line is illustrated with the example of the free energy surface behavior as a function of  pressure  at  $T =$ 2000 K (see Fig. \ref{fgr:FES} (left)). At the lower pressure, the LDL phase is more likely while at the  high pressure  HDL prevails.  The location at which they are equally probable defines the pressure where the two liquids coexist at the selected temperature. This behavior is consistent with a first-order transition, and the locus of such coexistence points defines the boundary line between the two phases. 

This behavior is to be contrasted with that at  $T =$ 2800 K, where pressure induces a continuous change from a LDL-like to a HDL-like structure (see Fig. \ref{fgr:FES} (right)). In this region of the phase diagram one is above the critical point.  However, also in this part of the phase diagram a thermodynamically significant line can be drawn. This is the so called Widom line defined as the locus of a point where the compressibility is the highest. At T=2800 K this point occurs at P=0.2 Pa as a consequence of the fact that the free energy surface curvature is very low and therefore the compressibility is large.

\begin{figure}[!ht]
  \centering
  \includegraphics[width=0.48\textwidth]{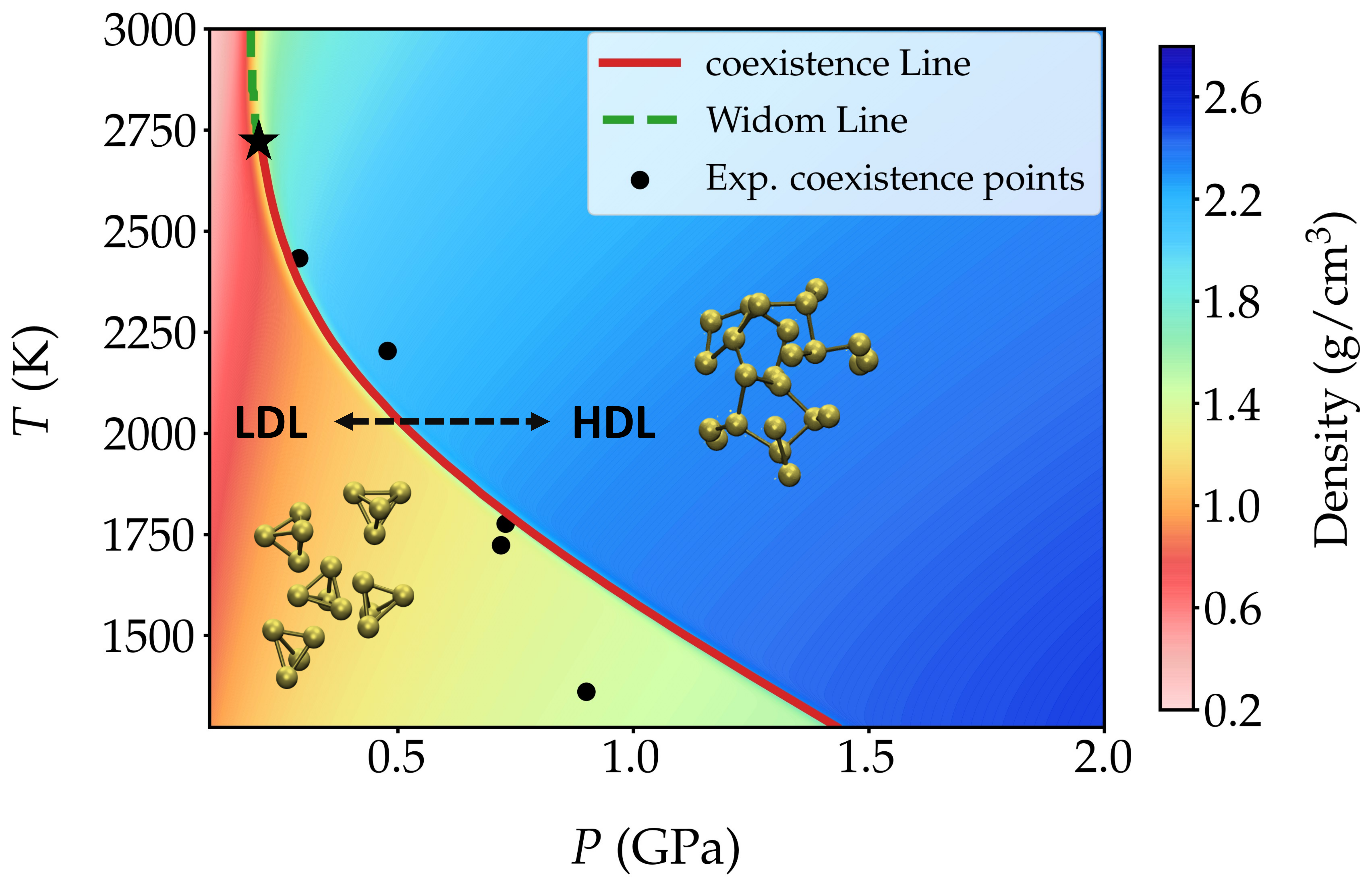}
  \caption{Liquid phosphorus phase diagram: The value of the density as a function of $T$ and $P$ is also plotted. The LL coexistence line is shown in red line. Black  points denote the experimental coexistence points. The calculated  LL critical point is indicated by the black star, and the Widom line (discussed also in section S4.1 in SI) is shown in dashed green line. Typical structures of the molecular and polymeric phases are depicted in ball-and-stick.}
  \label{fgr:Phase_transition}
\end{figure}

This analysis can be repeated at all desired  temperatures and pressures in the pressigned  MultiTP range and leads to the phase diagram in Fig. \ref{fgr:Phase_transition}. In agreement with experiments, the coexistence line has a negative slope. It can be clearly seen that there is a  density jump across the phase transition line. As the system approaches the critical point (whose position will be determined below) the jump eventually vanishes. In the same picture we also show two typical structural units of LDL and HDL. In LDL, the structural units are $P_4$ molecules while the HDL phase is characterized by the presence of small irregular branched multimer of different sizes that continuously form and break. One such multimer is shown in the picture. As we approach the critical point from the LDL side we see (see Fig. S9 in SI) that more and more P$_4$ units are broken. This is to be contrasted with what happens on the HDL side where the multimeric structure is maintained until one gets rather close to the critical point.

\begin{figure}[!h]
  \centering
  \includegraphics[width=0.4\textwidth]{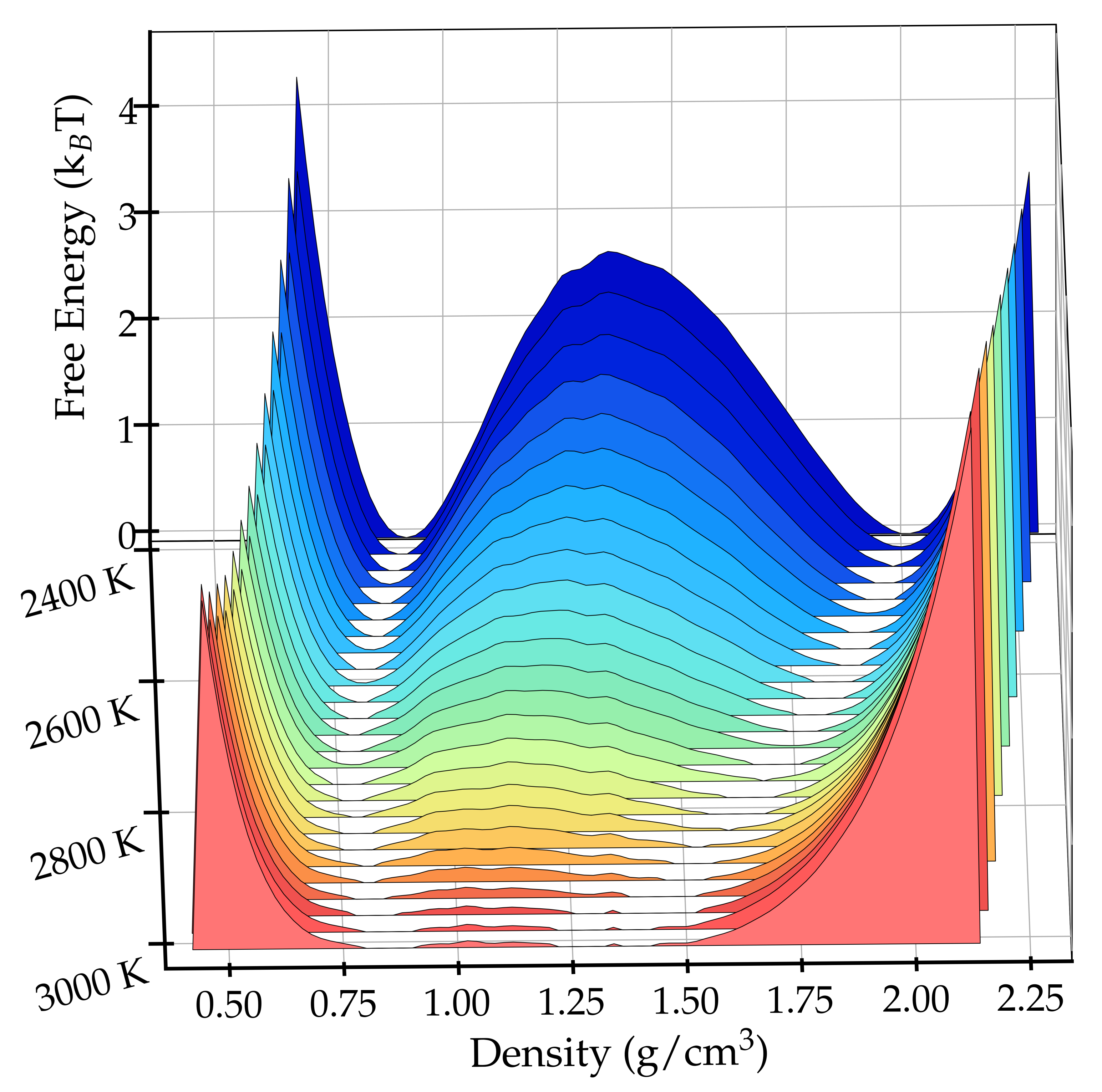}
  \caption{FESs as a function of the density and temperatures along the coexistence line. The color of the FESs varies from blue at $T =$ 2400 K to red at $T = $ 3000 K.}
  \label{fgr:finite}
\end{figure}

We now focus on the transition region. To this effect, we show in Fig. \ref{fgr:finite} the free energy profiles evolution along the coexistence curve. At the lower temperatures the LDL and HDL minima are well separated reflecting the presence of a first-order transition. However, as the temperature is increased, the barrier between the two phases vanishes and the two phases can no longer be distinguished. The barrier-less transition between the two regimes identifies the critical point. However, reading its precise  position from Fig. \ref{fgr:finite} is difficult and in addition one expects large finite size effects. For this reason we made a Binder cumulant analysis (see section S4.2 in SI) \cite{PhysRevLett.47.693,watanabe2012phase} that allows computing the thermodynamic value of the critical point. In such a way we  estimate for the critical temperature and pressure the values  $T_c \sim$ 2690 K and  $P_c \sim$ 0.2 GPa (see Fig. S7 in SI), that are in line with the indirect one suggested by the experimentalists \cite{Monaco2003}.

It is fascinating  to study the evolution  of the electronic structure as reflected in the electronic density of stated (DOS) and the inverse participation ratio (see section S4.3 in SI). We perform this analysis computing the electronic structure on selected configurations along the coexistence line.

 \begin{figure}[!hb]
  \centering
  \includegraphics[width=.48\textwidth]{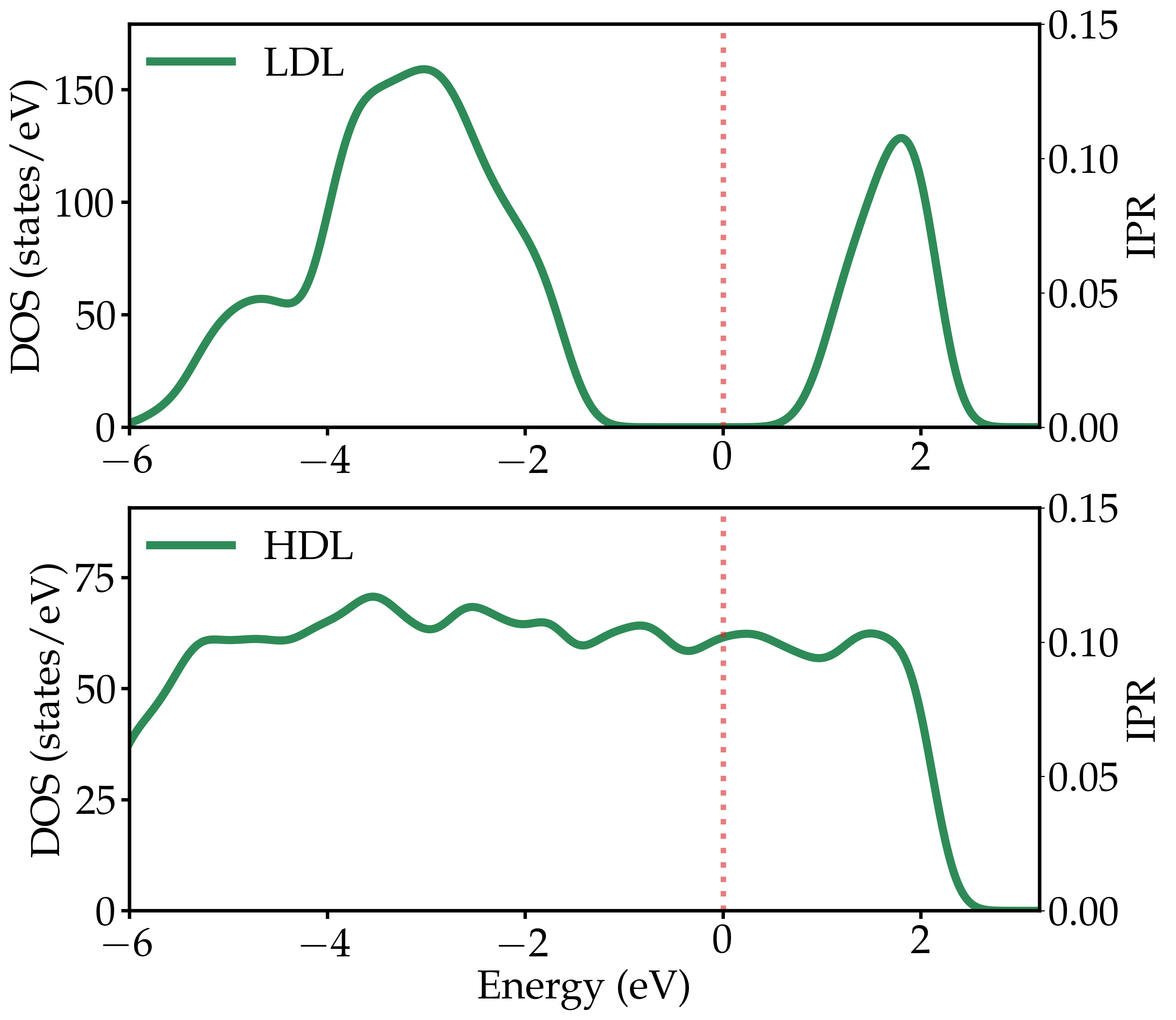}
  \caption{DOS for configurations in the LDL (top) and HDL (bottom) phase at $T = $1273 K. The Fermi level is shown in red dashed line.}
  \label{fgr:LDL_HDL_DOS}
\end{figure}

\begin{figure}[!h]
  \centering
  \includegraphics[width=0.48\textwidth]{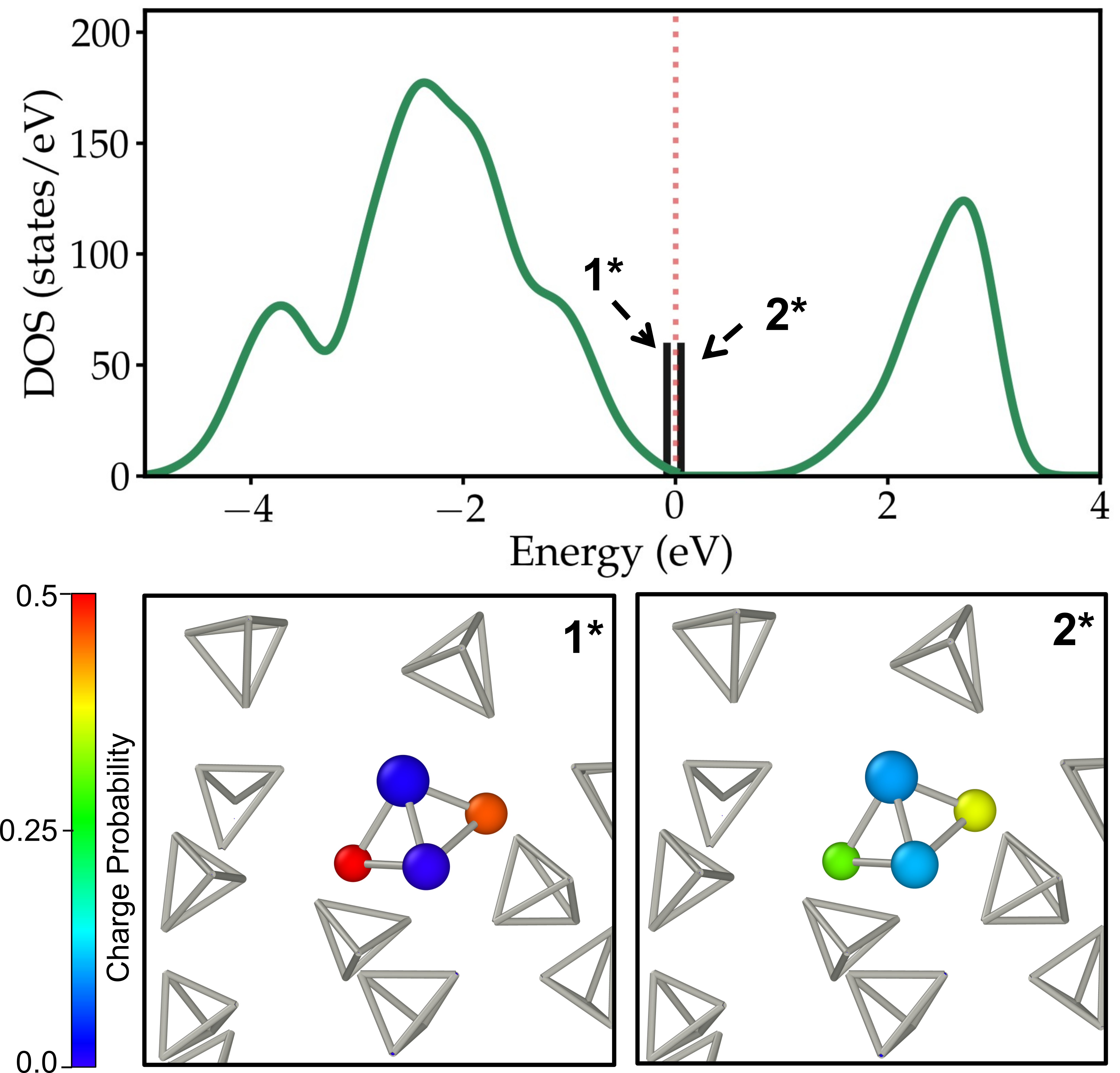}
  \caption{Top: DOS of a defective LDL configuration. The state 1$^*$ and 2$^*$ are localised on a P$_4$ molecule in which one of the bonds is broken. The Fermi level is shown in red dashed line. The probability density on the unbroken P$_4$, represented by stick only, is negligible, while it is significantly large on the broken P$_4$, that is represented in (bottom) ball-and-stick.}
  \label{fgr:DOS}
\end{figure}

Deep into the two-phase region, the LDL DOS clearly shows a gap (see Fig. \ref{fgr:LDL_HDL_DOS} (top)) and the occupied states are well localized as to be expected from a molecular liquid. In contrast, the HDL DOS has a metallic character (see Fig. \ref{fgr:LDL_HDL_DOS} (bottom)) and the electronic states are rather delocalized (see SI). The behavior of the LDL as it  approaches $T_c$ is rather intriguing since localized states appear in the bandgap. These states are localized on the partially broken P$_4$ molecules (Fig. \ref{fgr:DOS}), thus they will not contribute to a metal type conductivity. As we further approach $T_c$ the number of broken P$_4$ molecules increases and so does the number of states in the bandgap (see Fig. S9 in SI). Right at $T_c$, the electronic structure undergoes huge fluctuations and goes from a metallic like character to an insulator one (see Fig. S10 in SI), passing \emph{via} intermediate states such as those described above. 

Finally, we hope that this fascinating behavior can be experimentally probed since this is an unusual critical system in which a metal non metal transition is strongly coupled to a structural one. The success of our strategy that combines MultiTP-OPES with machine learning encourages us to study in the future even more complex systems.

We thank Michele Invernizzi, Chang Woo Myung, Luigi Bonati, and Omar Abou El Kheir for their help and the many useful discussions. The research was supported by the European Union Grant No. ERC-2014-AdG-670227/VARMET. We also thank the NCCR MARVEL, funded by the Swiss National Science Foundation. The computational resources were obtained from ETH Zurich Euler supercomputer.

\clearpage
\newpage
\bibliographystyle{unsrt} 
\bibliography{ref-ms}

\end{document}


\tableofcontents

\section{Computational details}
\subsection{System setup and equilibration}
The initial structures of the LDL phase was obtained by placing randomly 32 P$_4$ molecules (128 phosphorus atoms) in a periodically repeated cubic box.  In order to remove any nonphysical contacts, we optimized this initial structure. Subsequently, we carried out AIMD NVT simulation at $T$ = 1273 K followed by a NPT equilibration  at $P$ = 0.85 GPa. In order to obtain an initial HDL, a LDL configuration was brought to $T$ = 1500 K and $P$ = 5 GPa. The structure thus obtained was equilbrated following a schedule similar to that applied to the LDL phase. 

\subsection{AIMD simulations}
All AIMD simulations (NVT and NPT) were run using the CP2K software \cite{maintz2016lobster}. In these runs, the Perdew–Burke–Ernzerh (PBE) exchange-correlation density functional was used \cite{PBE}. The Kohn and Sham orbitals were expanded in a m-DZVP Gaussian basis while the plane wave expansion of the electronic density was truncated at an energy cutoff of 300 Ry. The core electrons were treated using the Goedecker-Teter-Hutter (GTH) pseudopotentials \cite{goedecker1996separable,hartwigsen1998relativistic} optimized for PBE. To deal with the metallic  configurations, we adopted a Fermi Dirac smearing of the occupation number of 0.2585 eV around the Fermi energy. To reduce the computational cost, only the $\Gamma$-point was used to sample the supercell Brillouin zone.

A time step of 2.0 fs was used in all AIMD simulations. Temperatures and pressures  were controlled using  Nos\'{e}-Hoover thermostat\cite{evans1985nose} and a  Nos\'{e}-Hoover-like barostat \cite{melchionna1993hoover} with coupling constants of 0.05 ps and 0.5 ps, respectively.

\subsection{DFT energy and force calculations}
\begin{figure}
  \centering
  \includegraphics[width=.8\textwidth]{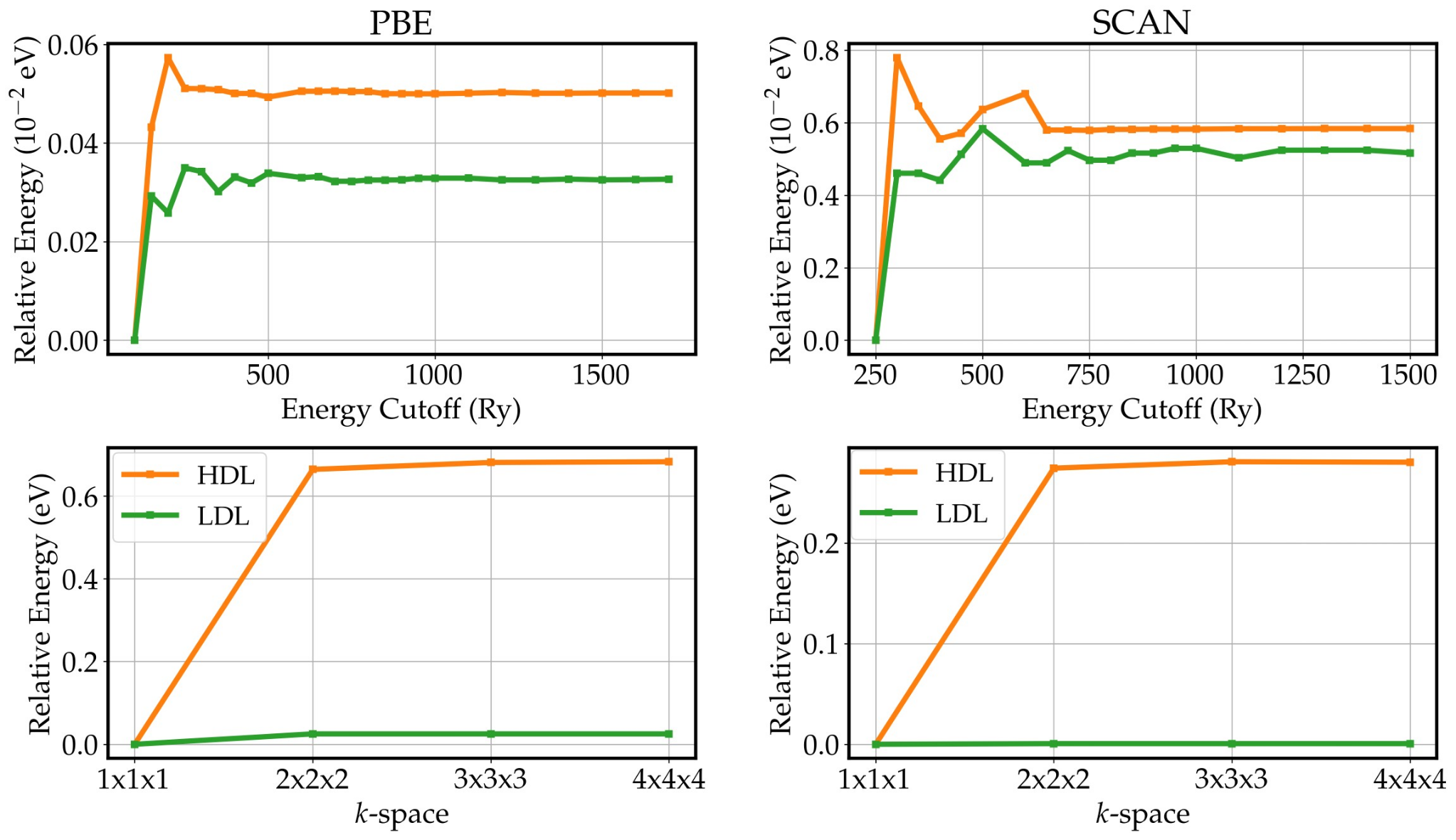}
  \caption{The convergence test for the energy cutoff and the $k$-space with the PBE and SCAN functionals}
  \label{fgr:Energy_cutoff_K_space}
\end{figure}

Single point energies and forces were calculated using two exchange-correlation density functionals, PBE \cite{PBE} and Strongly Constrained and Appropriately Normed (SCAN) \cite{SCAN}, and their two variants obtained by adding a D3 dispersion corrections \cite{D3-grimme2011effect}. Specifically, four different DFT methos, PBE, PBE+D3,Scan, Scan+D3, were used. In all cases, the core electrons were described using the GTH pseudopotential\cite{goedecker1996separable,hartwigsen1998relativistic} adapted to the different exchange-correlation functionals used. The smearing parameter is same as that of the AIMD simulations. However, the energy cutoff for PBE and SCAN-based calculations were increased to 400 and 700 Ry, respectively. Additionally, we used $k$-points grids of 2$\times$2$\times$2. These setups provide more accurate results (see Fig. \ref{fgr:Energy_cutoff_K_space}).

\newpage
\subsection{MuiltTP-OPES simulations}
In the MultiTP-OPES simulation one runs an appropriately biased simulation at reference temperature $T_0$ and pressure $P_0$. The bias is crafted such that after reweighing the properties of the system in a whole range of temperatures $(T_{min} < T_0 < T_{max})$ and pressures $(P_{min} < T_0 < P_{max})$ can be recovered. Details of this procedure can be found in Ref. \citenum{invernizzi2020rethinking} and \citenum{invernizzi2020unified}. 
  
In addition, since we want to drive transitions from one phase to the other, we enhanced the fluctuations of the intens ity of the first peak ($q=1.2$ {\AA$^{-1}$} ) of the Debye structure factor (Fig. \ref{eq:Debye}) that has contrasting values in the two liquid  phases, and it was used here as a CV. The Debye structure factor is defined as:
\begin{align}
    S({q}) = \frac{1}{N}\sum_{i=1}^N \sum_{j=1}^N\frac{sin (q\cdot{r_{ij}})}{q\cdot{r_{ij}}}
    \label{eq:Debye}
\end{align}
where, ${q}$ is the modulus of scattering wave vector, and $r_{ij}$ are the pair-wise distances. 

\begin{figure}[!h]
  \centering
  \includegraphics[width=.8\textwidth]{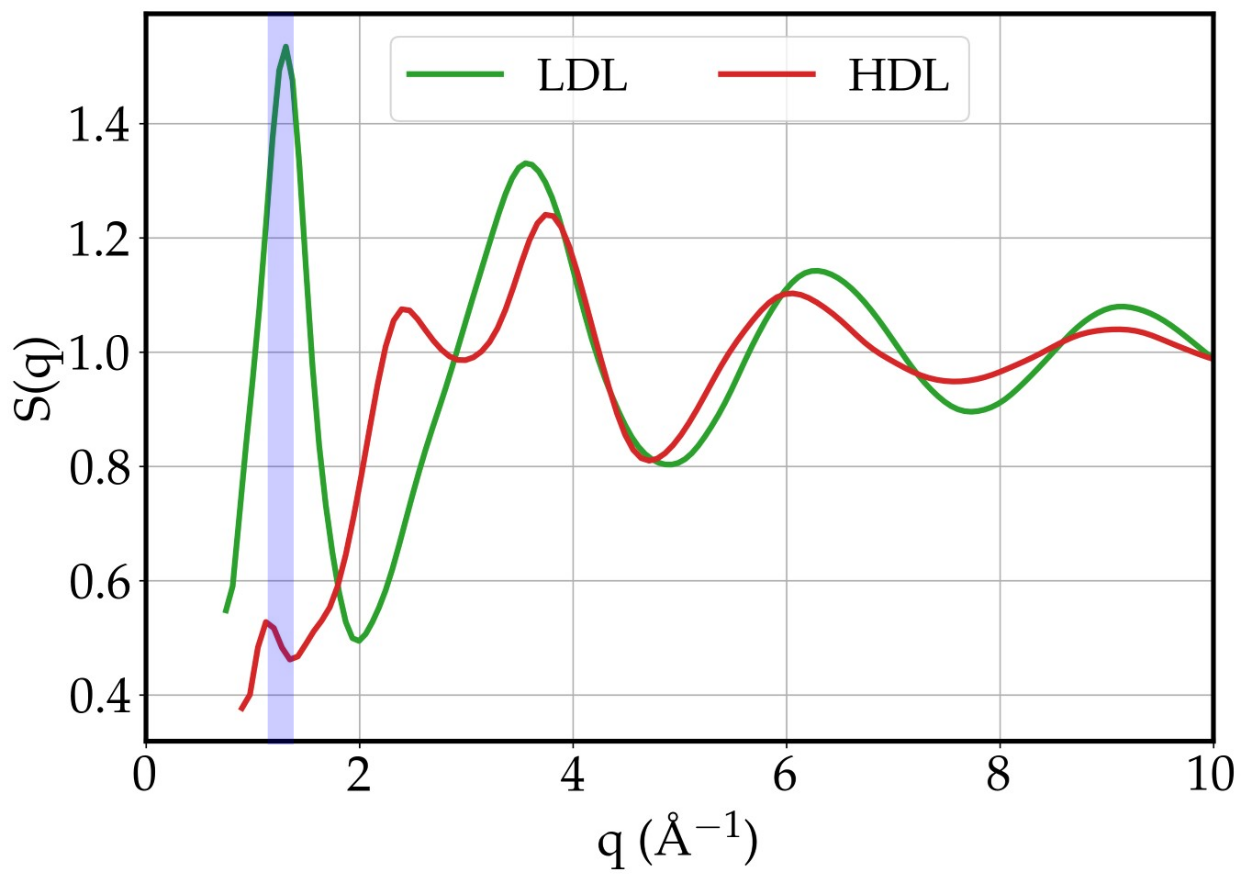}
  \caption{Debye structure factor profiles for the LDL and HDL phases. The transparent blue slab marks the first peak whose intensity was used as a CV for the MultiTP-OPES simulations.}
  \label{fgr:StructureF}
\end{figure}

The MultiTP-OPES simulations were run by patching the DeepMD-kit software \cite{wang2018deepmd} implemented in LAMMPS \cite{lammps} into a development version of PLUMED \cite{plumed}. Temperature and pressure  were controlled using the Nos\'e-Hoover thermostat \cite{evans1985nose} and the Parrinello-Rahman barostat \cite{parrinello1981polymorphic} with the relaxation times of 0.1 ps and 0.5 ps, respectively. A time step of 1.0 fs was used in all calculations.

In this work, we run MultiTP-OPES simulations in two cases. The first one is used in the DP-GEN iteration steps to sample the atomic configuration space (see section S2.1). The second one is utilized in the production simulations using the final NN models to study the phase diagram of liquid phosphorus (discussed in Section S3). 

\subsection{DeepMD setup}
The NN potentials were trained using the DeepPot-SE (Deep Potential-Smooth Edition) model \cite{end-to-end} with the DeePMD-kit package \cite {wang2018deepmd}. The cutoff radius smoothly decays from 5.5 {\AA} to 6.5 {\AA}. We used three hidden layers with (25, 50, 100) nodes/layer for the embedding network and four hidden layers with (240, 240, 240, 240) nodes/layer for the fitting network. The learning rate decays from 1.0 $\times$ 10$^{-3}$ to 5.0 $\times$ 10$^{-8}$. The batch size was set to 1. The prefactors of the energy and the force terms in the loss function change from 0.1 to 1 and from 1000 to 1, respectively. For the PBE and PBE+D3 NN models, we have included the virial term with prefactor value changing from 0.02 to 0.2 in the loss function,.

\section{NN potential training and validation}
\subsection{Collecting training set configurations}
\begin{figure}[!ht]
  \centering
  \includegraphics[width=.6\textwidth]{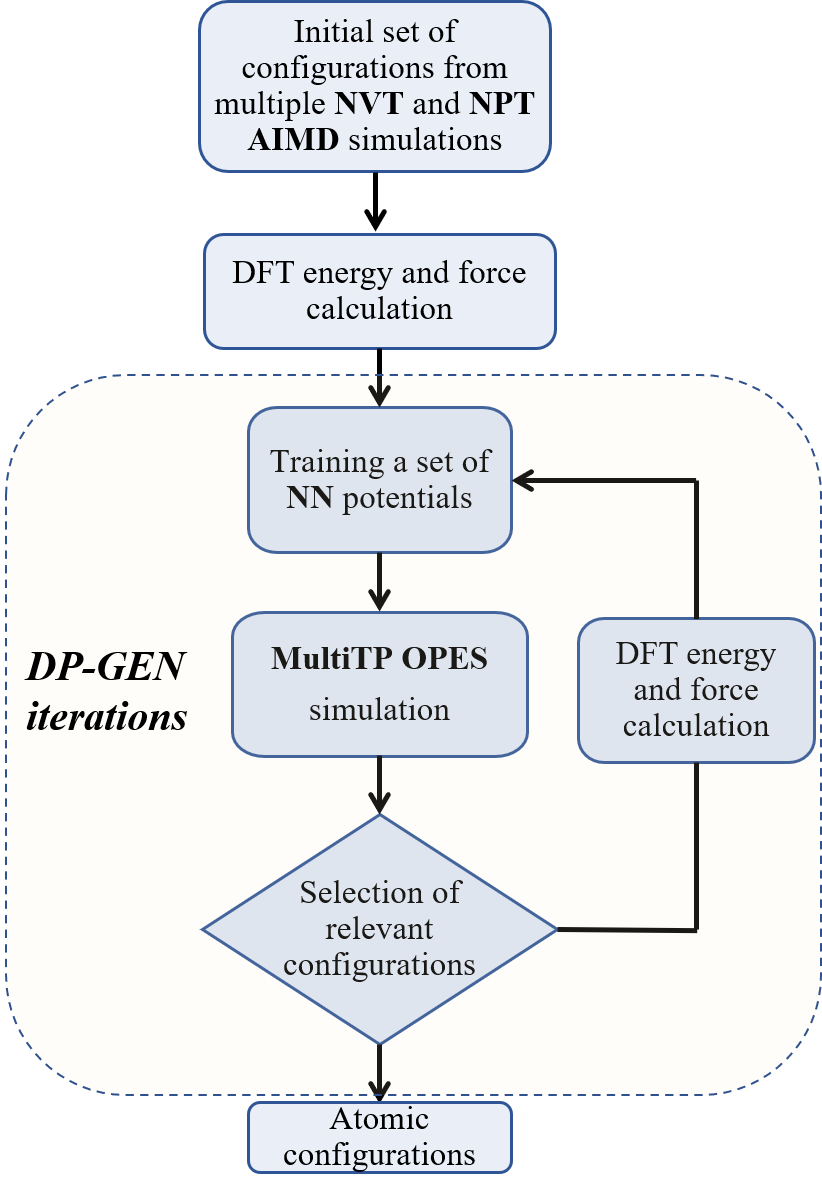}
  \caption{Flowchart for collecting configurations of the training set.}
  \label{fgr:flowchart}
\end{figure}

We have used an active learning procedure augmented with MultiTP-OPES method. This strategy, implemented within the DP-GEN package \cite{zhang2020dp}, allows us to collect configurations spanning a large area of the phase diagram. The active learning flowchart is shown in Fig. \ref{fgr:flowchart}. In what follows are the details of the active learning procedure.

\begin{figure}[!h]
  \centering
  \includegraphics[width=.8\textwidth]{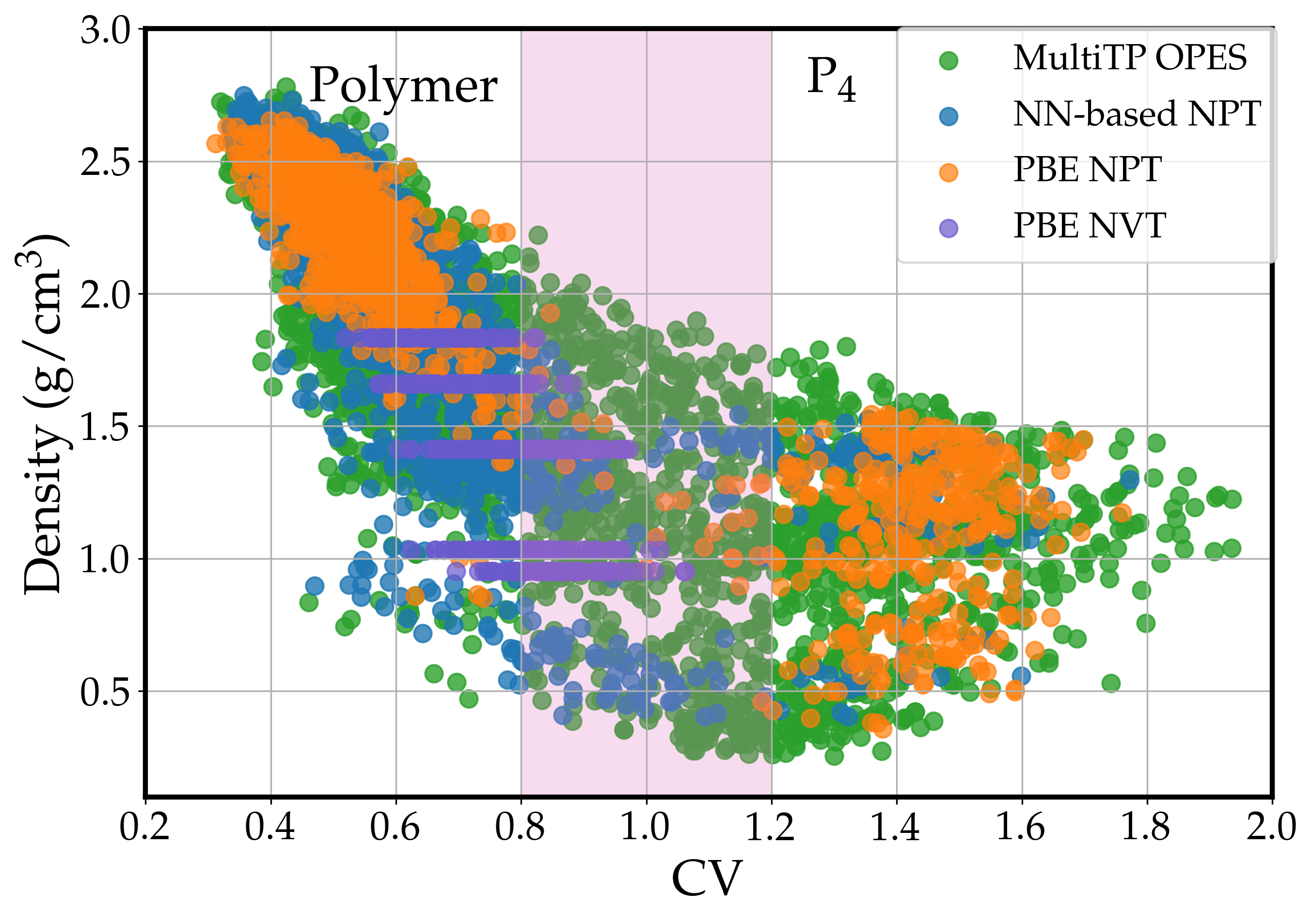}
  \caption{Training set configurations in the CV-Density space: green, blue, orange, and purple points correspond to the configurations sampled \emph{via} MultiTP-OPES, NN-based NPT, PBE NPT, and PBE NVT simulations, respectively. The lower the value of the CV, the more polymer-like a configuration is. The transition state configurations are the ones that have CV range from 0.8 to 1.2 (red shadow marked region).}
  \label{fgr:Trainingset_CV_Den_space}
\end{figure}

First we collected configurations from series of AIMD NPT and NVT simulations (see Section S1.2). The NPT simulations were performed in a range of temperatures from 1273 K to 3000 K and  pressures from 0.1 GPa to 2.0 GPa. We also performed NVT simulations at temperatures ranging from 1273 K to 3000 K starting from the HDL phase with varying box lengths from 15.5 to 19.0 {\AA} that corresponds to the density range 1.77 to 0.96 g/cm$^3$; these simulations provided some information on the phase transition region as shown in Fig. \ref{fgr:Trainingset_CV_Den_space}. In total, 32 NPT and 24 NVT simulations were performed. The  simulation time in the different runs ranged from 2 to 10 ps. Finally, we have $\sim$22700 atomic configurations in the initial training set (see Fig. \ref{fgr:Trainingset_CV_Den_space}). Subsequently, their PBE energies and forces were calculated.

These configurations, and the relative DFT energies and forces are the inputs of the DP-GEN iterations (see Fig. \ref{fgr:flowchart}). Each DP-GEN iteration involves following four steps:
\begin{itemize}
    \item {\bf Step 1:} Train four NN potentials using the same training set but different initial weights (training steps = 1.2 $\times$ 10$^6$);
    \item {\bf Step 2:} Run MultiTP-OPES simulations with the updated NN potentials to explore atomic configuration space;
    \item {\bf Step 3:} Select new relevant configurations based on a selection criteria (discussed later); 
    \item {\bf Step 4:} Calculate atomic energies and forces for the selected configurations using PBE method. Add these configurations, energies, and forces to the training set;
\end{itemize}

The DP-GEN iterations are continued until proper exploration of the configuration space of our interest is achieved, and the errors in the NN models are minimized (discussed later). 

\vspace{20pt}

We found that NN potentials trained only on AIMD (NVT and NPT) configuration were not stable, and the MultiTP OPES simulations run with these NN potentials crashed within a very short simulation time. This is due to the fact that the AIMD configurations were mainly focused on the limited regions of the polymer and $P_4$ basins and with insufficient configurations in the phase transition region. So for Step 2, in the initial four iterations, we used multiple NPT simulations with NN potentials to enlarge the configuration space. After obtaining slightly more stable NN potentials, we switch to MultiTP-OPES simulations to explore the whole phase diagram configuration space (See Fig. \ref{fgr:Trainingset_CV_Den_space}). The sampled configurations span a temperature range of 1273 K to 3000 K and pressure range from 0.1 GPa to 5.0 GPa.

The criteria for selecting relevant configurations in Step 3 is based on the model deviation that is defined by the maximal standard deviation of atomic forces predicted by four different NN potentials that were generated in Step 1.
The structures thus generated are labeled according to their model deviations. In our calculations, the structures with model deviations in the range of $[$ 0.2, 0.35 $]$ eV/{\AA} were labeled as candidate configurations. The lower bound of the model deviation was set to a value that is slightly higher than the average model deviation of the initial training set. The structures with model deviation below the lower bound are already well-represented in the training set. On the other hand, the choice of the upper bound was based on a rule ([0.15 $\sim$ 0.30 ev/{\AA}] + lower bound) suggested in Ref. \citenum{zhang2020dp}. The configurations above the upper bound are non-physical and thus not included in the training set. In each DP-GEN iteration, we run OPES simulations at different thermodynamics conditions. At each thermodynamics condition, we select a maximum of $N_{max} = 250$ and a minimum of $N_{min} = 30$ out of many allowed labeled candidate structures as done in Ref. \citenum{YANG2021} . 

The criteria to stop the DP-GEN iterations is as follows. In each DP-GEN iteration, we monitored the percentage of candidate configurations. Lower the number of newly candidate configurations better is the chance of converged exploration of the configuration space. In our case, once the percentage of candidate configurations reaches $\sim$ 10 \% and remains almost unaltered for another few iterations, we exit the loop. This made sure that the configuration space is properly explored. In this way, we collected about 74000 atomic configurations.

\clearpage

\clearpage
\subsection{Training NN models}
With the atomic configurations, energies and forces collected in the previous section, we trained the final PBE-based NN model with long training steps (6.0 $\times$ 10$^6$). To investigate the effect of different density functionals on the phosphorus NN potential, we obtained three additional NN models that were trained on the same set of configurations but energies and forces calculated using PBE+D3, SCAN, and SCAN+D3 density functionals. Finally, we obtain four NN potentials that are labeled as PBE, PBE+D3, SCAN, and SCAN+D3 models (see Table S1).

\subsection{Evaluation of NN models}
Mean absolute errors (MAEs) of the final four different NN models prediction in terms of the atomic energies and forces on the test sets are given in Table S1. Configurations in each test set were extracted from the production MultiTP-OPES simulations (discussed in section S3) which were performed using the corresponding NN potentials. For the system containing $N =$ 128 phosphorus atoms, each test set sampled $\sim$ 8000 configurations in a range of temperatures 1273 K $< T <$ 3000 K and pressures 0.1 GPa $< P <$ 2.0 GPa. For the system of $N =$ 256, $\sim$ 800 configurations sampled in a range of temperatures 2400 K $< T <$ 3000 K and pressures 0.1 GPa $< P <$ 0.4 GPa were included. 

\begin{figure}[!ht]
  \centering
  \includegraphics[width=0.7\textwidth]{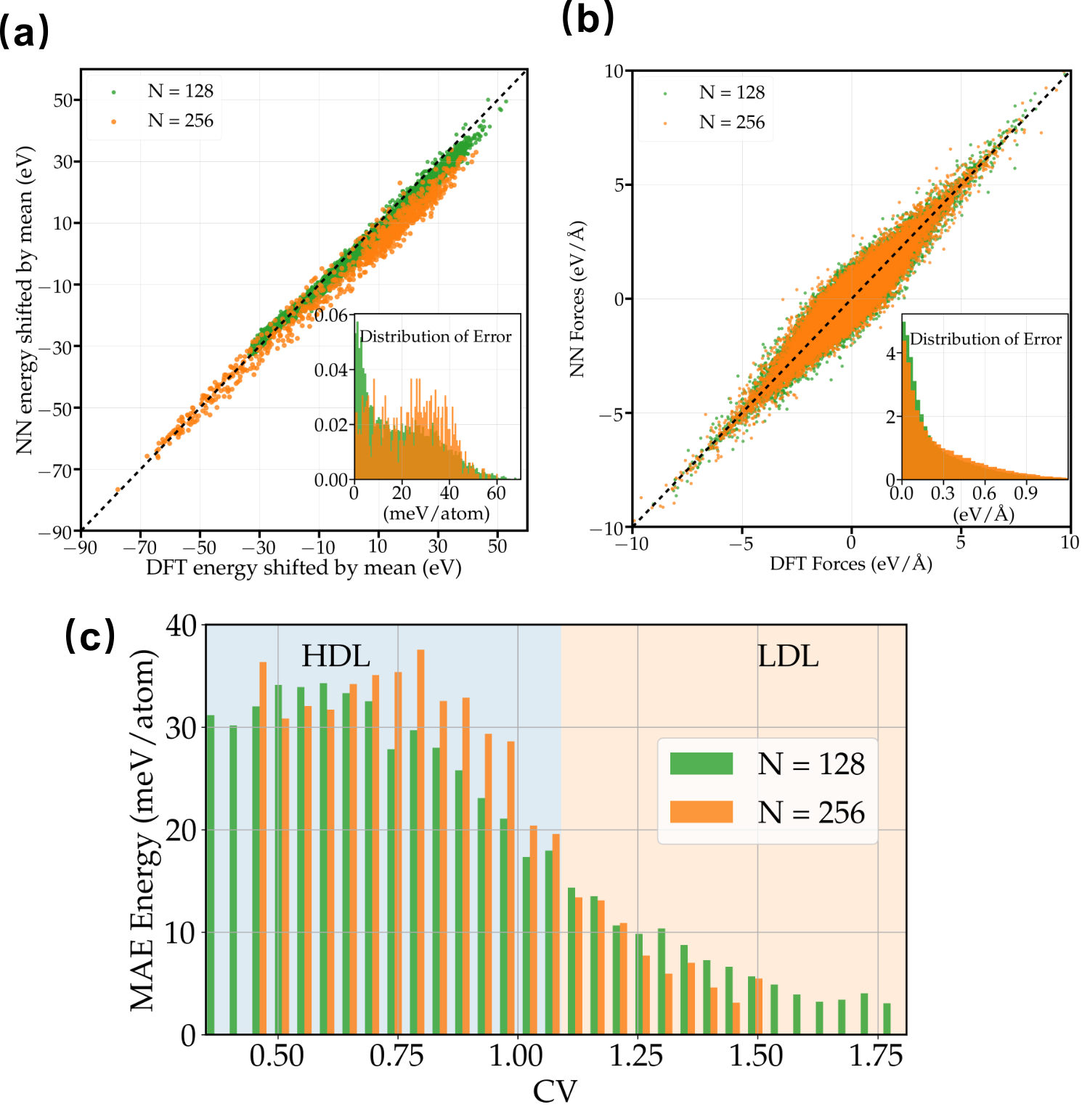}
  \caption{Comparison of atomic energies (a) and forces (b) on test sets calculated by SCAN$+$D3 DFT method and its corresponding NN model. The configurations in test sets were extracted from the production MultiTP-OPES simulations. Two different systems consisting of $N=$ 128 (green) and $N=$ 256 (orange) phosphorus atoms were considered. Energies in panel (a) are shifted by the mean value of the DFT energies. Insets in panel (a) and (b) illustrate the probability distributions of the absolute difference in energy and force between the DFT and NN model; (c) The distribution of the MAEs of the energies as a function of the structure factor CV.}
  \label{fgr:NN_vallidation}
\end{figure}

The comparison of SCAN$+$D3 DFT and the corresponding NN predicted atomic energies and forces over the test sets is given in Fig. \ref{fgr:NN_vallidation}.  Note that the MAEs of energies in the HDL phase are much higher than that of the LDL phase. We anticipate that the larger MEAs in HDL phase is due the complex coordination environment of the P atoms in the polymeric phase. A similar observation was also reported by Deringer \emph{et al.} \cite{Deringer2021}.

Table S1: The MAEs in terms of the atomic energies and forces on the test sets for four different NN models.
\hspace{20pt}
  \label{Tab:NNmodels}
\begin{center}
 \begin{threeparttable}
 \begin{tabular}{ccccc} 
  \hline
  \hline
  \multirow{1}{*}{Models}
         & \multicolumn{2}{c}{MAE of Energy (meV)} &\multicolumn{2}{c}{MAE of Force (meV/\AA)} \\
  \hline
  PBE      & & 23.0  & & 261  \\
  PBE+D3   & & 18.4  & & 248 \\
  SCAN     & & 20.8  & & 240  \\
  SCAN+D3  & & 19.8 (24.4)$^a$ & &  227 (283)$^a$ \\
  \hline
  \hline

 \end{tabular}
  \begin{tablenotes}
   \footnotesize
   \item $^a$System that contains $N=$ 256 phosphorus atoms.
  \end{tablenotes}
\end{threeparttable}
\end{center}

\clearpage
\section{MultiTP OPES simulations}
\begin{figure}[!h]
  \centering
  \includegraphics[width=0.8\textwidth]{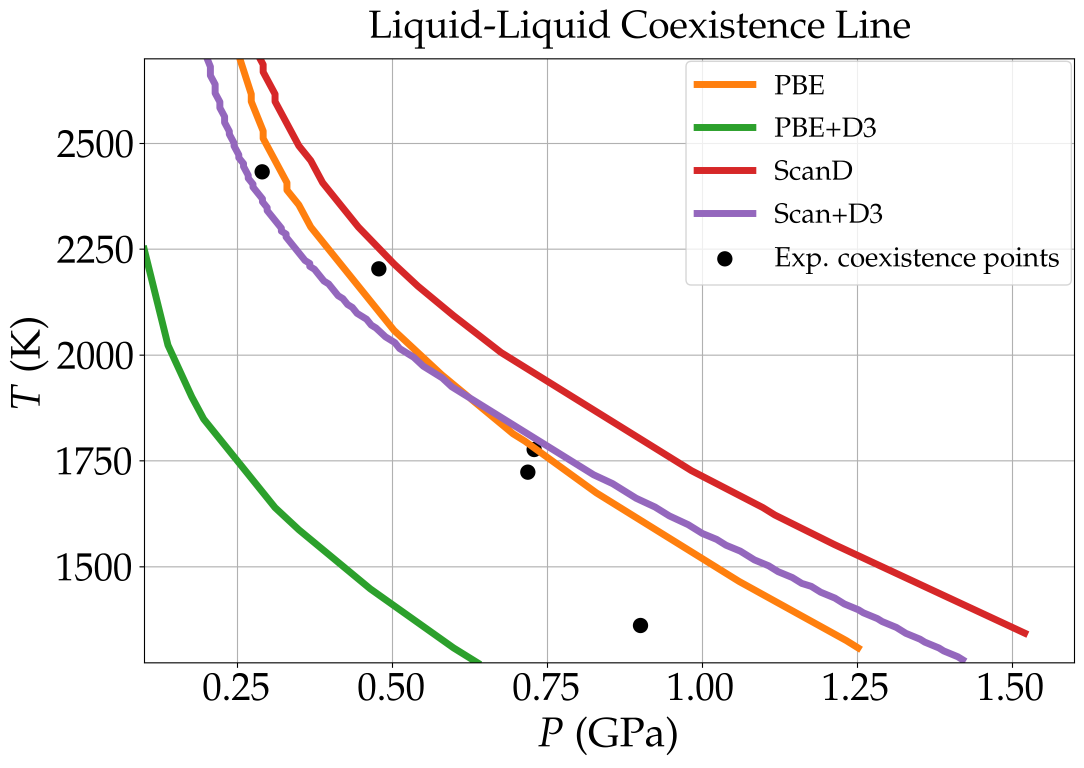}
  \caption{Comparison of the liquid-liquid coexistence line of phosphorus calculated by MultiTP-OPES with different NN models (PBE, PBE+D3, SCAN, and SCAN+D3). Black points denote the  experimental coexistence points.}
  \label{fgr:MultiTP_LLTL_all}
\end{figure}

To explore the whole liquid-liquid phase diagram, we run MultiTP OPES simulations with 128 phosphorus atoms at $T_{min} = 1273 < T_0 = 2200 < T_{max} = 3000$ K and $P_{min} = 0.1 < P_0 = 0.3 < P_{max} = 2.0 $ GPa. Four independent MultiTP OPES simulations with the four NN models (PBE, PBE+D3, SCAN, and SCAN+D3) were carried out. After reweighing these simulations, we obtained the liquid-liquid coexistence lines that are shown in Fig. \ref{fgr:MultiTP_LLTL_all}. It is clear that the PBE and SCAN+D3 based NN models best approximate the experimental coexistence points. 

We carried out additional MultiTP OPES simulations with the SCAN+D3 NN model mainly focusing on the critical region to locate the critical point. In these simulations, we considered a temperature range 2400 to 3000 K and pressure range 0.1 to 0.4 GPa with the reference temperature $T_0$ = 2600 K and pressure $P_0$ = 0.15 GPa. The calculated free energy surfaces are presented in Fig. 3 of the main text. 

As we have mentioned in the main text, the free energy barrier height between the LDL and HDL phases depends on the system size. For this reason, we performed additional simulations with larger systems containing 256 and 512 phosphorus atoms focusing on the critical region. These simulations were utilized to calculate the system-size dependent Binder parameter as discussed in Section S4.2.  

In both sets of MultiTP-OPES simulations, we have used four walkers that share the same bias. The simulations length ranges from 75 to 100 ns.

\section{Additional results}
\subsection{Widom line}
We calculated the Widom line as the locus of the compressibility maxima $K_T^{max}$ for $ T \geq T_c$. To locate the Widom line, first we calculated the compressibility $K_T$ from the fluctuations of volume $V$ (NPT simulations) using the equation:
\begin{align}
    K_T = \frac{<V^2>-<V>^2}{<V>k_BT}
\end{align}
where $<V>$ is the ensemble average of $V$ and $k_B$ is the Boltzman constant. 

For every $T$ and $P$ points in the predefined TP range, the corresponding $K_T$ can be obtained by reweighing the MultiTP-OPES simulations. At each $T$, we plotted the $K_T$ as a function of $P$. Projecting the maximum $K_T$ of these profiles ($K_T^{max}$) on the TP plane provides the Widom line as shown in Fig. 2 of the main text.

\subsection{Binder parameters}
We located the critical point using the Binder parameter ($U_N$). In the scaling limit, $U_N$ for systems of different number of particles $N$ is expected to intersect at a common point, that corresponds to the critical point. The $U_N$ is calculated using the following equation,
\begin{align}
    U_N = 1- \frac{<\rho^4>_N}{3<\rho^2>^2_N}
\end{align}
where, $<\rho^k>_{N}  = \int{(\rho-\bar{\rho})^k}{\mathbf{P}_N(\rho)} {d\rho}, k=2, 4$ is the $k$-th moment of the density, $\rho$. The $\bar{\rho}$ and ${\mathbf{P}_N(\rho)}$ in the integrand are the expectation value and the distribution function of $\rho$.

We obtained the values of $U_N(T,P)$ as a function of $T$ and $P$ on the phase diagram by reweighing the MultiTP OPES simulations. Then, we extracted the values of $U_N(T,P)$ along those $T$ and $P$ in which the compressibility is maximum $K_T^{max}$. 

\begin{figure}
  \centering
  \includegraphics[width=0.7\textwidth]{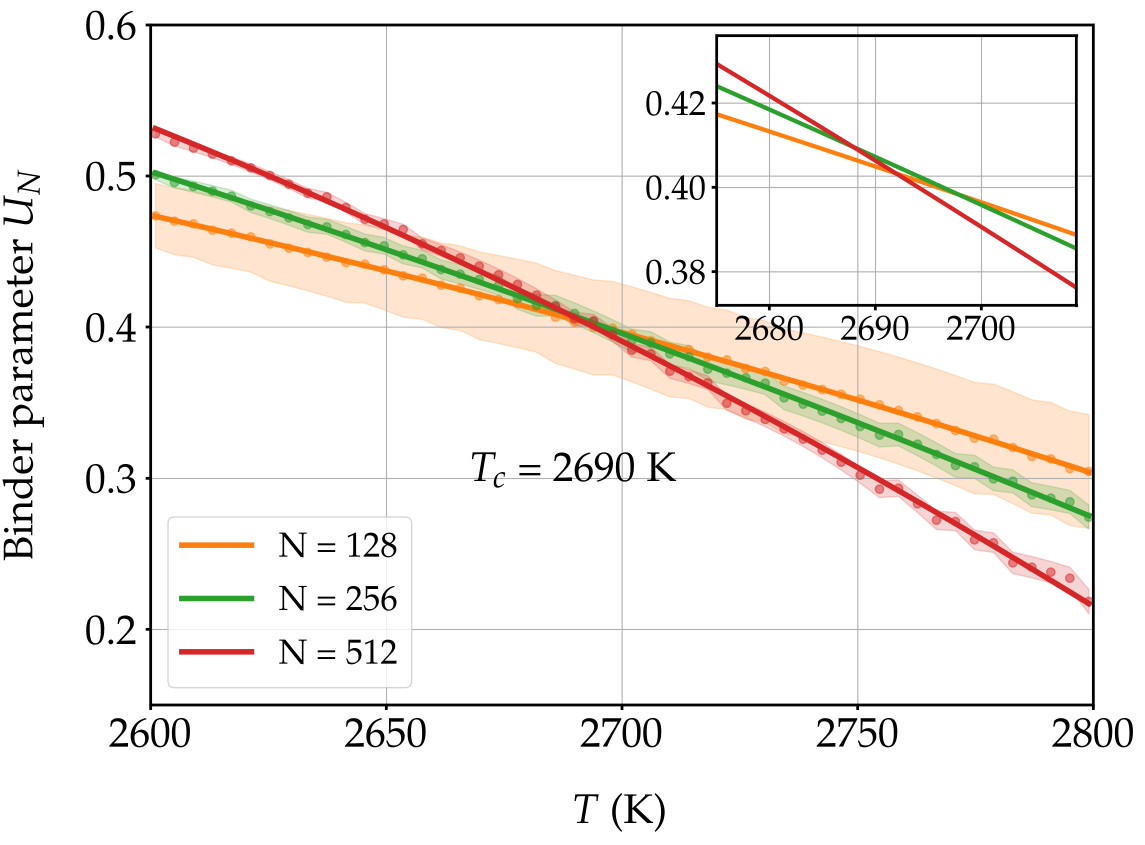}
  \caption{Binder parameter $U_N$ as a function of the temperature $T$, for different system sizes $N$ = 128, 256, and 512. The solid lines are the second degree polynomial fitting of the raw data. The shaded areas indicate the statistical error calculated using the weighted block-average technique (number of blocks = 4) discussed in Ref. \citenum{invernizzi2020unified}.}
  \label{fgr:Binder_line}
\end{figure}

Following this strategy, we calculated the $U_N$ for the three systems with N = 128, 256, and 512 phosphorus atoms. From Fig. \ref{fgr:Binder_line} we can see that these three curves approximately intersect at one point that corresponds to the critical point $T_c \sim$ 2690 K and $P_c \sim$ 0.2 GPa.

\subsection{Density of states and inverse participation ratio}
We studied the electronic property of liquid phosphorus by calculating the electronic density of states (DOS) for configurations in HDL and LDL phases as well as those along the coexistence line. Furthermore, to quantify the localisation of the electronic states, we calculated the inverse participation ratio (IPR). The IPR for the $n$th eigenstate ${\psi}_n$ is given as,

\begin{align}
    IPR(\psi_n) = \frac{\sum_{i=1}^{N}{|{\phi}_{n,i}|^4}}{(\sum_{i=1}^{N}{|{\phi}_{n,i}|^2})^2}\
 =\sum_{i=1}^{N}|q_{n,i}|^2
\end{align}

where ${\phi}_{n,i}$ is the projected coefficient of the $i$th basis set orbital in the $n$th Kohn–Sham eigenstate ${\psi}_n$. The $q_{n,i}$ is the probability of finding the (normalized) eigenstate ${\psi}_n$ in basis set orbital $i$. If the eigenstate is ideally localised at one particular basis orbital $i$, the corresponding IPR is equal to one. On the contrary, the IPR for the equally delocalised state is equal to $1/N$, namely, a higher IPR signifies a higher degrees of localization. 

To calculate the DOS and IPR, we first run a self-consistent calculation using a $k$-points grid of 2 $\times$ 2 $\times$ 2 to generate electron density, and then performed another non-self-consistent calculation with a $k$-points grid of 4 $\times$ 4 $\times$ 4. All simulations were performed with the Quantum Espresso 6.6 package \cite{QE-2009,QE-2017,QE-3}. We used a cubic system consisting of 128 phosphorus atoms, PBE exchange-correlation functional, and the PAW pseudopotential. we set the plane-wave cutoff as 40 Rydberg (Ry) and the DOS were calculated using Gaussian smearing with broadening at 0.3 eV.

 \begin{figure}
  \centering
  \includegraphics[width=.9\textwidth]{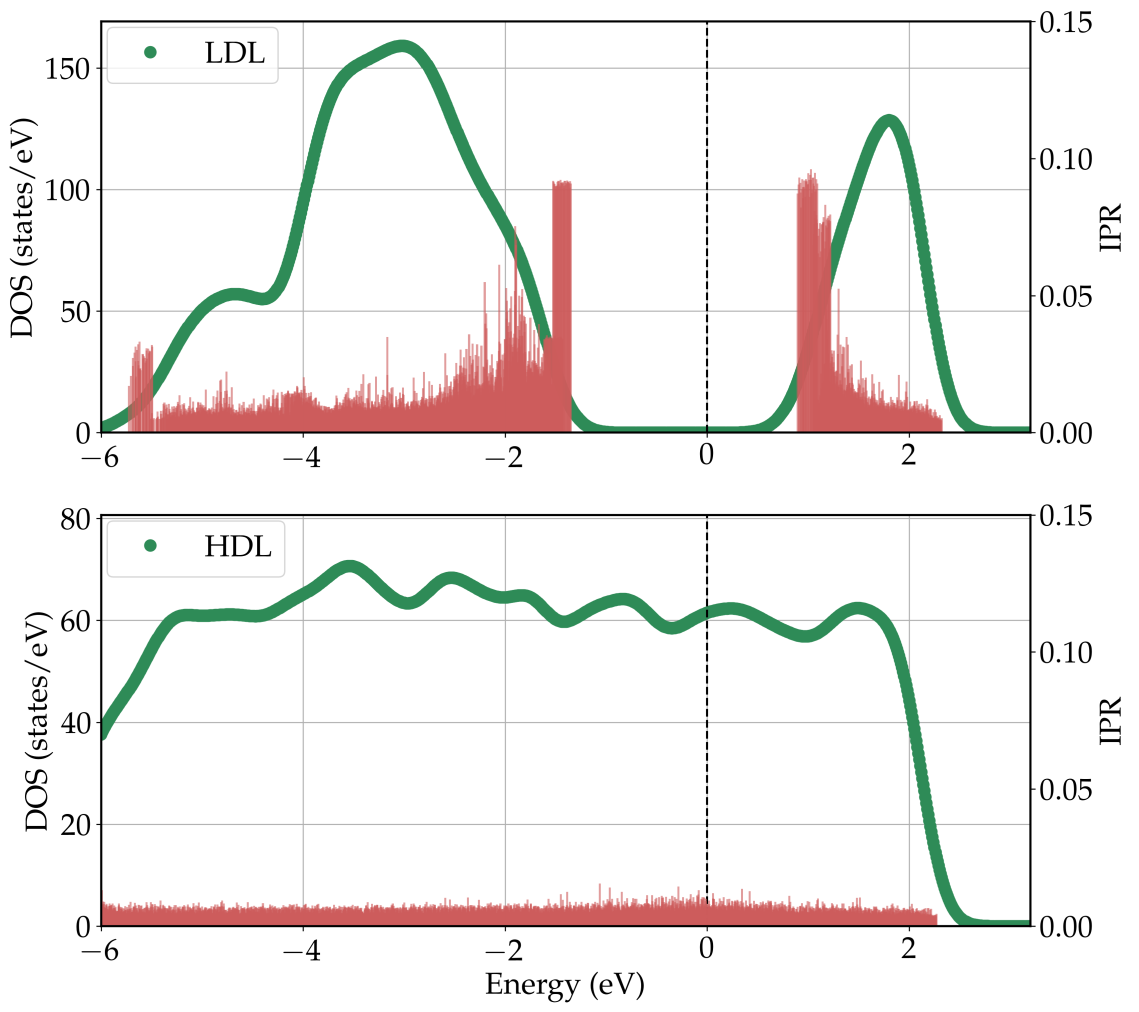}
  \caption{The total DOS (Green line) and IPR (Red spikes) for configurations in the LDL (top) and HDL(bottom) phase at $T = $1273 K. The Fermi level is shown in black dashed line.}
  \label{fgr:LDL_HDL_DOS}
\end{figure}

\clearpage
\subsection{Temperature-dependented structural evolution}
\begin{figure}
  \centering
  \includegraphics[width=1.0\textwidth]{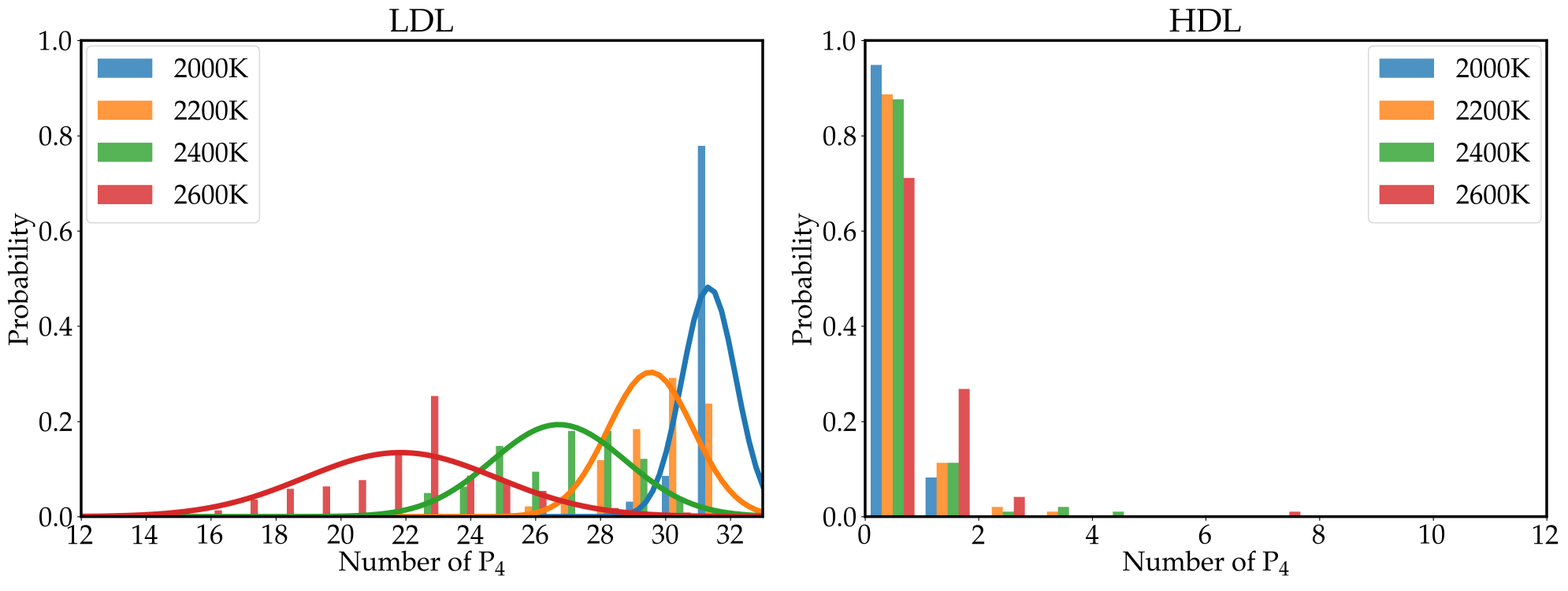}
  \caption{The histogram of the number of P$_4$ units in each structure as a function of $T$ when we approach the critical point from both the LDL (left) and HDL (right) side. For each thermodynamic condition, 100 structures extracted from 2 ns standard NPT trajectory were analyzed. (For the LDL phase, we increase the $T$ from 2000 K to 2600 K at $P =$ 0.15 GPa. Instead, for the HDL phase, the simulations were carried out at various $P$: 0.53 GPa, 0.45 GPa, 0.35 GPa, and 0.30 GPa.)}
  \label{fgr:Count_p4}
\end{figure}

To quantify the structural evolution of the LDL and HDL phase as they approach the critical region, we performed a few standard NPT simulations at different $T$ and $P$. The number of P$_4$ molecules is used as a marker. Ideally, the LDL phase should contain 32 P$_4$ molecular units. As we increase the $T$ from 2000 K to 2600 K the P$_4$ molecules break resulting in a dramatic reduction of the P$_4$ count (Fig. \ref{fgr:Count_p4}). On the other hand, the HDL phase retains its polymeric structure until it reaches to 2600 K where a few P$_4$ molecules appear.

\clearpage
\subsection{Structural evolution in the critical region}
\begin{figure}[!h]
  \centering
  \includegraphics[width=0.86\textwidth]{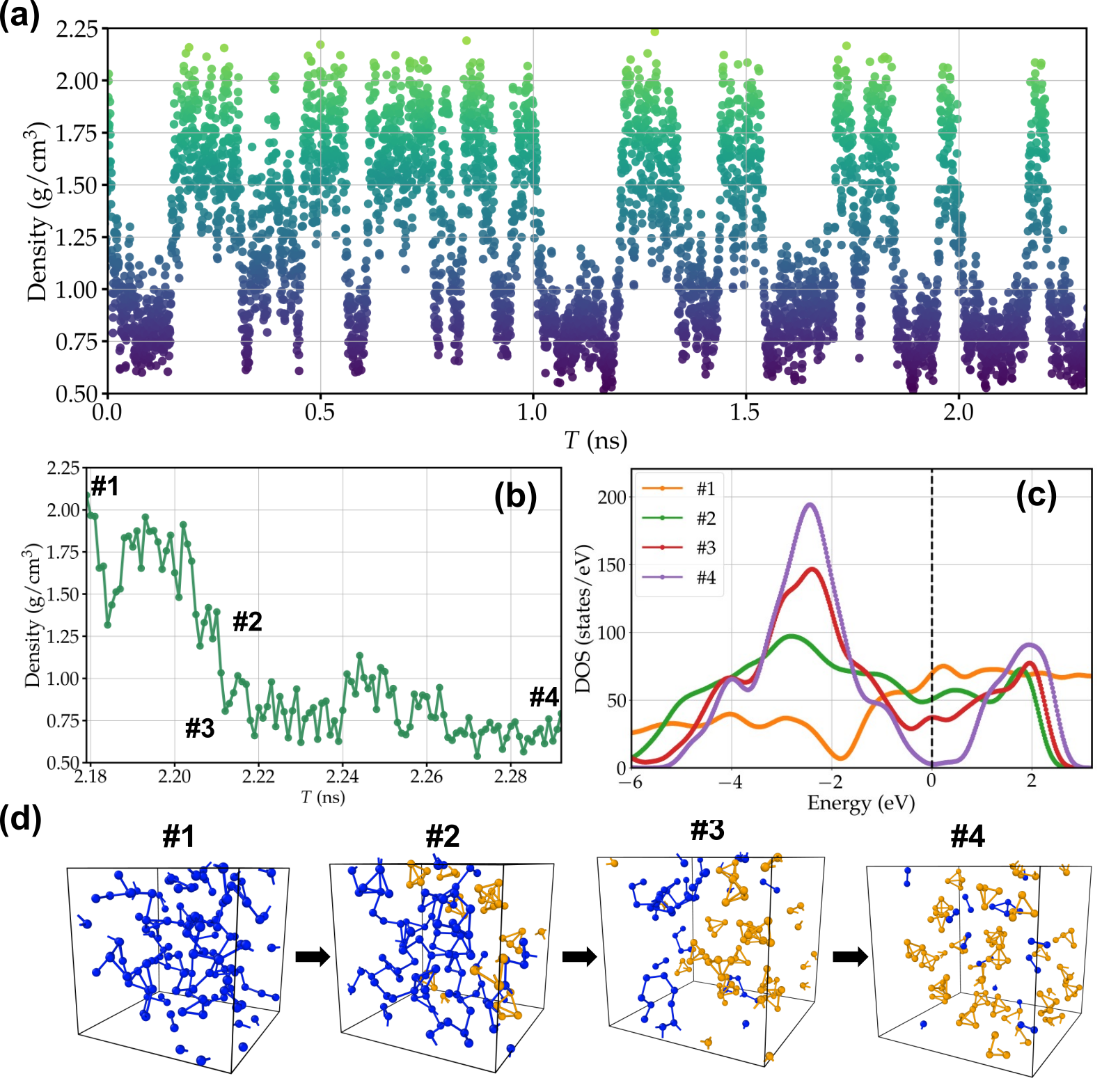}
  \caption{(a) Density as a function of the simulation time obtained from an unbiased NPT simulation run at the $T$ = 2700 K and $P$ = 0.2 GPa; The evolution of (b) density, (c) DOS, and (d) structures as the system transits from the polymeric to the molecular phase. Atoms belong to polymers are shown in blue while those belong to the P$_4$ units are displayed in orange.}
  \label{fgr:DOS_change}
\end{figure}

Furthermore, we investigated the structural evolution of the system in the critical region. For this purpose, we run an additional unbiased NPT simulation at $T$ = 2700 K and $P$ = 0.2 GPa. Fig. \ref{fgr:DOS_change}(a) shows large density fluctuations which is expected for a system in its critical state. Fig. \ref{fgr:DOS_change}(b) shows a typical density evolution as the system transits from the polymeric to molecular phase. 
The density change accompanies with the transformation from the metallic to the non-metallic phase as revealed by the DOS (Fig. \ref{fgr:DOS_change}(c)). The structural evolution going from the polymeric phase (\#1) to the molecular phase (\#4) via the intermediates (\#2 and \#3) is illustrated in Fig. \ref{fgr:DOS_change}(d).

\bibliography{ref-si}